\documentclass[11pt]{article}

\usepackage[utf8]{inputenc}
\usepackage[T1]{fontenc}
\usepackage{lmodern}
\usepackage[margin=1in]{geometry}
\usepackage{graphicx}
\usepackage{booktabs}
\usepackage{amsmath,amssymb}
\usepackage{xcolor}
\usepackage{listings}
\usepackage{caption}
\usepackage{microtype}
\usepackage[hidelinks]{hyperref}
\usepackage{authblk}

\graphicspath{{figs/}}

\definecolor{codebg}{rgb}{0.97,0.97,0.97}
\title{\bfseries Operating Multi-Node Full Fine-Tuning on NVIDIA B300:\\[2pt]
A Field Report on Telemetry-Based Triage, Negative Results, and Operational Hardening}

\author[1]{Seon Ho Kim}
\author[1]{Ui Jeong Jeon}
\author[1]{Su Hyeon Kim}
\author[1]{Min Tae Hwang}
\affil[1]{Samsung SDS\\ \texttt{\{seonho1499.kim,\ uijeong.jeon,\ sensible.kim,\ mr\_rye.hwang\}@samsung.com}}
\date{\today}

\begin{document}
\maketitle

\begin{abstract}
We report operational experience full-fine-tuning a 32.76B-parameter dense model
(Qwen3-32B) on 16 $\times$ NVIDIA B300 (two nodes, FSDP / ZeRO-3)---among the first
published field accounts on this accelerator. \textbf{We claim no new algorithm.}
The individual mechanisms we use are established practice; our contribution is the
\emph{integrated} field experience and a set of \emph{calibrated measurements} on
new hardware. Concretely we offer four practitioner artifacts. \textbf{(1)} A
\textbf{B300-calibrated power-draw triage table} that distinguishes compute /
communication / data-starvation / checkpoint-or-deadlock / idle by board wattage,
motivated by the well-known failure of GPU-\emph{utilization\%} to reflect real
progress (it reads 100\% during an NCCL hang). \textbf{(2)} A set of \textbf{honest
negative results} that dispel common optimization folklore at this scale: a
controlled A/B in which per-step NFS reading matches a pretokenized local cache
(${\sim}53$k tok/s) because the corpus fits in page cache and the job is
compute-bound; and a reconstruction of an earlier ``throughput collapse'' as
NFS/CPU contention rather than a storage-medium limit. \textbf{(3)} \textbf{Calibrated
$4/8/16$-GPU strong-scaling and GPU$\cdot$hour numbers} on B300 (near-linear, as
expected in this regime; we report absolute values as reference data).
\textbf{(4)} A \textbf{worked failure case}---an epoch-end NCCL deadlock from
per-rank token-packing imbalance---together with a 2.7-second pre-run invariant
gate and an external watcher that turn multi-hour silent failures into instant
rejections. We are explicit that this deadlock and its remedy correspond to
PyTorch's documented \texttt{Join} context manager and the standard
\texttt{drop\_last} / equalize-to-minimum practice; we position our in-situ
instantiation against that prior art and report the GPU-hours the failure cost and
the gate saves. The transferable takeaway is operational, not algorithmic:
\emph{for data-dependent data-parallel jobs, watch power rather than utilization,
and verify invariants before launch---a passing smoke test is not evidence of a
safe full run.}
\end{abstract}

\section{Introduction}

This is an experience report. We do not introduce a new training algorithm, a new
parallelism strategy, or a new optimizer. Instead we share what it actually took
to operate a multi-node full fine-tuning job on \textbf{NVIDIA B300}---brand-new
hardware (Blackwell Ultra, 2025--26) for which little operational experience is yet
public---and the measurements and triage practices we found valuable and believe
transfer to other practitioners.

We are upfront about novelty, because it shapes how this report should be read.
The building blocks we rely on already exist and are well documented: FSDP/ZeRO-3
sharding; the fact that uneven per-rank inputs desynchronize collectives (the
documented motivation for PyTorch's \texttt{Join} context manager);
\texttt{drop\_last} / equalize-to-minimum batch handling; checkpoint/restart fault
tolerance; and the use of hardware telemetry to detect stalls. None of these is
our invention. What we offer is the \emph{integration} of these into a working
operational discipline on B300, plus \emph{calibrated numbers}---power bands,
strong-scaling efficiency, data-pipeline throughput, and the GPU-hour cost of a
real silent failure---that are otherwise hard to find for this hardware and
workload size.

The report is organized around four practitioner-facing deliverables:
\begin{itemize}
  \item \textbf{Telemetry-based triage} (\S\ref{sec:triage}): a B300-calibrated
  mapping from board \emph{power} (not utilization\%) to execution phase, with the
  flight recorder as confirmatory tool, packaged as a one-line monitor and a
  decision procedure.
  \item \textbf{Negative results} (\S\ref{sec:negative}): controlled measurements
  showing where we \emph{wasted} effort---local-cache optimization that bought no
  throughput at this scale---and a correction of an earlier misdiagnosis.
  \item \textbf{Calibrated scaling reference} (\S\ref{sec:scaling}): full-epoch
  $4/8/16$-GPU throughput and GPU$\cdot$hour numbers on B300, as reference data.
  \item \textbf{A worked failure $+$ operational hardening} (\S\ref{sec:case},
  \S\ref{sec:hardening}): an epoch-end deadlock from packing imbalance, positioned
  explicitly against \texttt{Join}/\texttt{drop\_last}, plus a 2.7-second preflight
  gate and external watcher, with a cost accounting.
\end{itemize}

The recurring theme is operational rather than algorithmic: in distributed
training, \emph{utilization lies, power tells the truth}, and \emph{a passing
smoke test is not evidence of a safe full run} when per-rank workload is
data-dependent.

\section{Background and Experimental Setup}

\subsection{Full fine-tuning memory and the need for sharding}
Full fine-tuning a dense model in bf16 mixed precision with Adam requires
${\sim}16$~bytes/parameter: parameters bf16 (2) $+$ gradients bf16 (2) $+$ fp32
Adam master (4) $+$ $m$ (4) $+$ $v$ (4). For Qwen3-32B (32.76B params) this is
\textbf{525~GB of model states}, far beyond a single GPU. We shard with
\textbf{PyTorch FSDP in \texttt{FULL\_SHARD} mode (equivalent to ZeRO-3)} across
all 16 GPUs, giving 32.8~GB of model states per GPU; the measured live footprint
including activations (gradient checkpointing off) is \textbf{82~GB/GPU},
${\sim}30\%$ of the B300's 275~GB.

\subsection{Cluster and software}
\begin{center}
\begin{tabular}{ll}
\toprule
Component & Value \\
\midrule
GPUs & 16 $\times$ NVIDIA B300 SXM6 AC (2 nodes $\times$ 8); 1100 W TGP, 275 GB HBM \\
Interconnect & InfiniBand (NDR) $+$ GPUDirect RDMA; NVLink intra-node \\
Software & NGC container (PyTorch 2.12, transformers 5.12, native FSDP) \\
Storage & shared NFS (\texttt{/dataset}); per-node local NVMe cache; ${\sim}4$~TB RAM/node \\
Model & Qwen3-32B (32.76B, dense) \\
Dataset & Nemotron-Personas-Korea, 1{,}000{,}000 records, ${\sim}1.01$B tok/epoch \\
\bottomrule
\end{tabular}
\end{center}

\subsection{Training configuration}
Sequence length 2048; micro-batch 1; gradient accumulation 8; world size 16,
giving a global batch of 128 sequences $=$ \textbf{262{,}144 tokens/step}. One
epoch is ${\sim}3{,}854$ steps. AdamW, cosine warmup, lr $1\mathrm{e}{-5}$, bf16
\texttt{MixedPrecision}, \texttt{Qwen3DecoderLayer} auto-wrap, gradient
checkpointing \textbf{off} (memory headroom makes recomputation pure waste; off is
${\sim}1.6\times$ faster---see \S\ref{sec:gradckpt}). Checkpoints are bf16
consolidated HF-format shards written to NFS every 500 steps, plus once at epoch
end.

\subsection{Data pipeline}
\label{sec:pipeline}
Records are sharded across ranks with \texttt{datasets.shard(world, rank)}
(disjoint per rank). Each record is tokenized once and \textbf{packed into fixed
2048-token blocks} persisted as a node-local NVMe \texttt{.i32} file
(${\sim}1.9$~GB/node). We adopted packing after a production run exhibited a
sustained \textbf{data-starvation collapse from 52k to 21k tok/s} (the
${\sim}575$~W ``uneven-utilization'' signature of \S\ref{sec:triage}), driven by
single-threaded tokenization plus page-cache cooling on the shared NFS under
multi-tenant load. A later \emph{controlled} measurement (\S\ref{sec:negative})
shows the storage \emph{medium} itself is not the bottleneck, which localizes the
residual value of packing to \textbf{determinism} (a bit-exact, resumable token
stream) and \textbf{robustness to NFS/CPU contention} rather than throughput. The
per-rank block count is $\lfloor \text{file\_bytes} / 4 / 2048 \rfloor$---the
quantity that turns out to be unequal across ranks and drives the failure case of
\S\ref{sec:case}.

\section{Telemetry-Based Triage: Watch Power, Not Utilization}
\label{sec:triage}

Our single most useful operational practice was to stop trusting GPU utilization
and start reading \textbf{board power}. We make no claim that power-based
diagnosis is new---vendor guides, SRE write-ups, and DCGM-based tooling all use
power and related telemetry to detect stalls (\S\ref{sec:related}). Our
contribution here is narrow and concrete: \textbf{the B300-calibrated wattage
bands}, and an integrated decision procedure that classified every stall we met
without attaching a profiler.

\subsection{Why utilization lies}
\texttt{nvidia-smi}'s utilization percentage reports only that \emph{a kernel is
scheduled}, not that it does useful compute. NCCL collectives, memory copies, and
IO waits all register as 100\% ``utilization'' while the GPU makes little
progress. The flagship symptom (\S\ref{sec:case}) is that during a full NCCL
deadlock utilization stays pinned at 100\% even though throughput is exactly zero.
Board power, by contrast, tracks real work: it falls when the silicon waits.

\subsection{The B300 power-draw bands}
\begin{center}\small
\begin{tabular}{lll}
\toprule
Power (per GPU) & Utilization pattern & Phase / interpretation \\
\midrule
\textbf{${\sim}940$--$1040$~W} & 100\%, uniform & compute (matmul-bound)---healthy \\
\textbf{${\sim}640$~W} & 100\%, uniform & communication; if sustained, overlap failing \\
\textbf{${\sim}575$~W} & \textbf{55--100\%, uneven} & data starvation (stragglers) \\
\textbf{${\sim}190$--$195$~W} & 1 rank 0\% / rest 100\% (deadlock) & hang / IO \\
 & \emph{or} rank 0 only 0\% (checkpoint) & (distinguish by shape) \\
\textbf{${\sim}135$~W} & 0\%, all ranks & idle \\
\bottomrule
\end{tabular}
\end{center}
Absolute wattages are hardware/firmware-specific (\S\ref{sec:threats}); the
\emph{ordering} and \emph{per-rank-shape discriminators} below are what transfer.

\begin{figure}[t]
\centering
\includegraphics[width=\linewidth]{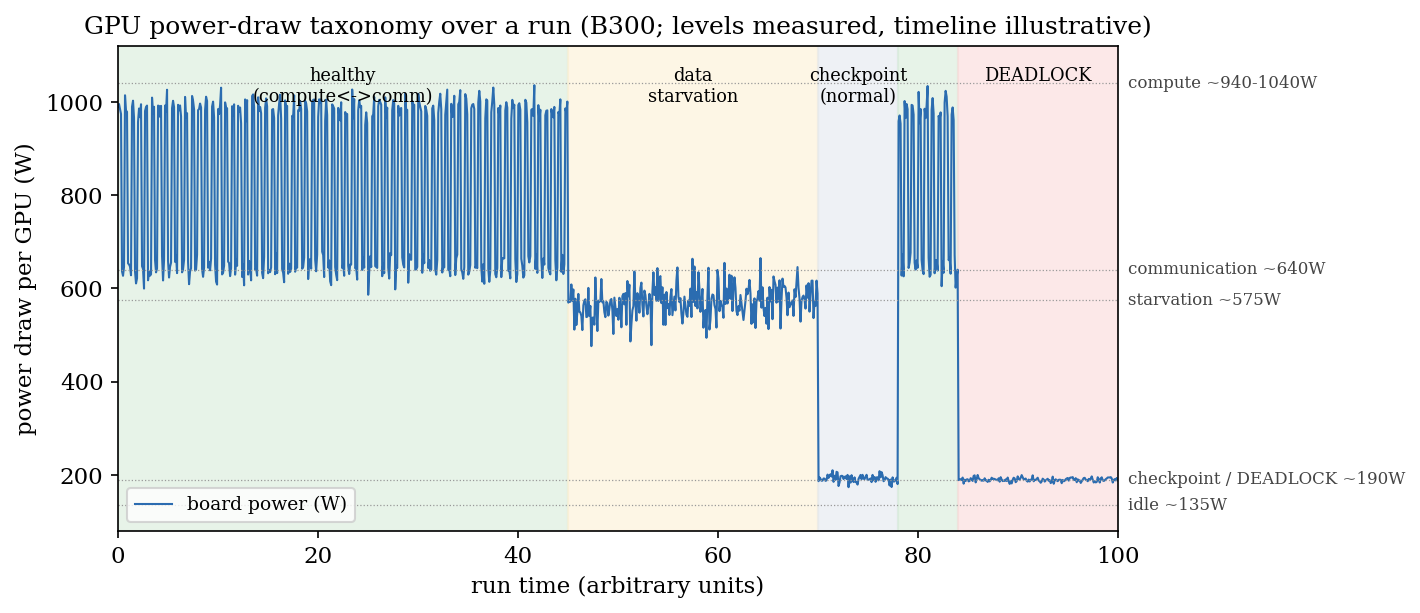}
\caption{GPU power-draw taxonomy over a run. Power levels are measured B300
values; the timeline is an illustrative arrangement of phases. Utilization can
read 100\% in several of these states, so power $+$ per-rank shape---not
utilization---is the reliable signal.}
\label{fig:power}
\end{figure}

\subsection{Two discriminations that resolve ambiguous states}
\begin{enumerate}
  \item \textbf{Deadlock vs.\ healthy checkpoint} (both ${\sim}190$~W). In a
  \emph{normal} checkpoint only \textbf{rank 0} drops to 0\% (it gathers shards and
  writes ${\sim}62$~GB to NFS) while others wait, resuming within ${\sim}2$~min. In
  a \emph{deadlock} the shape is asymmetric the other way---\textbf{one rank at 0\%
  and the rest pinned at 100\%} with power floored---and never recovers.
  \item \textbf{Starvation vs.\ communication} ($575$ vs.\ $640$~W). Communication
  keeps all ranks \emph{uniformly} busy; starvation produces rank-to-rank
  utilization scatter, because the bottleneck is upstream of the GPUs.
\end{enumerate}

\subsection{The triage procedure}
\label{sec:triageproc}
Combined, the bands and discriminators form a fast decision procedure needing only
\texttt{nvidia-smi} and, for confirmation, the NCCL flight recorder:
\begin{enumerate}
  \item Sample power $+$ per-rank utilization:
\begin{lstlisting}
nvidia-smi --query-gpu=index,utilization.gpu,power.draw,memory.used \
           --format=csv -l 1
\end{lstlisting}
  \item Map power to a band. ${\sim}940$--$1040$~W uniform $\to$ healthy; stop.
  \item ${\sim}190$~W $\to$ inspect per-rank shape: rank-0-only-0\% $\to$ normal
  checkpoint (wait ${\sim}2$~min); one-rank-0\%-rest-100\% persisting across two
  samples $\to$ suspect deadlock.
  \item Confirm with the flight recorder
  (\texttt{TORCH\_NCCL\_TRACE\_BUFFER\_SIZE}): different collectives on different
  ranks is conclusive (\S\ref{sec:case}).
\end{enumerate}
This classified the deadlock of \S\ref{sec:case} in seconds, before we opened the
flight recorder, and we wire its signature into the external watcher
(\S\ref{sec:hardening}).

\section{Negative Results: Where Optimization Effort Was Wasted}
\label{sec:negative}

Experience reports are most useful when they record what \emph{did not} work. We
present these precisely because we initially believed the opposite and spent effort
accordingly.

\subsection{A pretokenized local cache buys no throughput here}
We A/B-tested the two data paths at 16 GPUs, 300 steps each, holding everything
else fixed. \textbf{A} $=$ shared NFS, per-step \texttt{datasets.shard} read $+$
per-step tokenization (no packing). \textbf{B} $=$ pretokenized, packed,
node-local cache. To expose any cold penalty in A we dropped the page cache on both
nodes immediately before the cold run, then re-ran the identical rows warm.

\begin{center}
\begin{tabular}{lrr}
\toprule
Path & tok/s (median) & vs B \\
\midrule
A --- NFS per-step read $+$ tokenize, \textbf{cold} (post drop\_caches) & 52.8k & $1.00\times$ \\
A --- NFS per-step read $+$ tokenize, \textbf{warm} & 53.0k & $1.00\times$ \\
B --- pretokenized packed node-local cache & 53.0k & --- \\
\bottomrule
\end{tabular}
\end{center}

All three are within ${\sim}0.5\%$. The only slow step is the first one after a
cache drop ($33.9$k tok/s---first-touch $+$ JIT warmup), gone by step 10. The
reason is a $1000\times$ headroom: the corpus is ${\sim}4$~GB against
\textbf{4~TB of per-node RAM}, so after first touch it is resident in page cache;
and even a true cold miss is cheap---a 32B FSDP step takes ${\sim}5$~s while
reading 4~GB cold takes ${\sim}3$~s at the measured floor:

\begin{center}
\begin{tabular}{lrrl}
\toprule
Method & GB/s & Gbps & Bounds \\
\midrule
warm page cache (RAM) & 6.34 & 50.7 & memcpy ceiling \\
multi-stream concurrent O\_DIRECT (8/4 ranks) & 3.56 & 28.5 & N-rank concurrent supply \\
fio seq O\_DIRECT QD32$\times$4 & 3.54 & 28.4 & parallel drive peak \\
O\_DIRECT QD1 & 2.61 & 20.9 & single-stream drive floor \\
cold buffered (drop\_caches, full set) & 1.31 & 10.5 & true first-touch miss \\
\bottomrule
\end{tabular}
\end{center}

\textbf{Lesson and cost.} Before measuring, we treated NFS as a suspected
bottleneck and built the pretokenize-and-cache pipeline expecting a throughput
win. The win was zero at this scale---standard page-cache behavior once a dataset
fits in RAM, but not obvious to us in the moment. The cache is still worth keeping,
for \emph{determinism and contention-robustness} (\S\ref{sec:contention}), not
speed. \textbf{This conclusion inverts once the dataset exceeds RAM}---then every
epoch re-incurs cold reads and locality optimizations begin to dominate---so we
report it as scale-conditional.

\subsection{The ``52k\,$\to$\,21k collapse'' was contention, not the medium}
\label{sec:contention}
This seems to contradict an earlier incident in which throughput halved from 52k
to 21k tok/s mid-run. Reconciling the two localizes the real bottleneck. The
collapse occurred under \textbf{single-threaded tokenization}
(\texttt{TOKENIZERS\_PARALLELISM=false}, CPU load average ${\sim}15$) \textbf{and a
contended, cooling shared NFS}---not the quiet, parallel-tokenized conditions of
the A/B. With parallel tokenization and an uncontended fabric, even a fully cold
NFS read matches the local cache. The bottleneck was therefore
\textbf{tokenization-CPU throughput and NFS contention}, never the storage medium.
The earlier incident is real; our initial framing of it as a ``storage
bottleneck'' was imprecise, and we correct it here.

\subsection{Gradient checkpointing was the wrong default}
\label{sec:gradckpt}
With 82~GB live on a 275~GB GPU, activation memory was never the constraint, so
recomputation bought nothing and cost ${\sim}1.6\times$ throughput (33k vs.\ 53k
tok/s). ``Always enable gradient checkpointing'' is good advice when memory-bound;
in a memory-rich regime it is pure waste---verify before enabling.

\section{A Worked Failure Case: The Epoch-End Deadlock}
\label{sec:case}

We present the failure that consumed the most time, both as a concrete application
of the \S\ref{sec:triage} triage and as an honest case study. \textbf{Neither the
failure mode nor the fix is new}---both correspond to documented PyTorch behavior
(\S\ref{sec:priorart}). The value is the worked B300 diagnosis and the hardening it
motivated (\S\ref{sec:hardening}).

\subsection{Observed symptom}
A from-pretrained run progressed normally for ${\sim}5$ hours: loss
$1.60 \to 0.74 \to {\sim}0.68$, throughput ${\sim}53$k tok/s, all 16 GPUs at 100\%
utilization. At \textbf{step 3{,}850 of 3{,}857}---entering the end-of-epoch
checkpoint---all 16 GPUs hung for \textbf{three hours} until the NCCL watchdog
surfaced it. (This pre-fix run had no evenfix, so each rank ran toward its own
block count; evenfix later pins every rank to the global-minimum 3{,}854 steps,
\S\ref{sec:case}.) No \texttt{dmesg} OOM, no segfault, no error: pure silence, with
utilization still reading 100\%---the exact trap that motivated \S\ref{sec:triage}.

\subsection{Diagnosis via triage $+$ flight recorder}
Power had floored to ${\sim}190$~W with \textbf{one rank at 0\% and the rest at
100\%}---the deadlock shape, not the rank-0-only checkpoint shape. The NCCL flight
recorder dump was unambiguous:
\begin{itemize}
  \item \textbf{Rank 3} had entered an \texttt{all\_reduce}---the
  \texttt{dist.barrier()} guarding the end-of-epoch \texttt{save\_checkpoint}.
  \item \textbf{Ranks 0--2, 4--15} (fifteen) were blocked in a
  \texttt{reduce\_scatter}---the backward-pass gradient reduction, i.e.\ still
  training.
\end{itemize}
Two different collectives at the same wire slot; NCCL waits forever. The fabric was
innocent: throughput had been a steady 53k tok/s and IB bandwidth nominal.

\begin{figure}[t]
\centering
\includegraphics[width=0.92\linewidth]{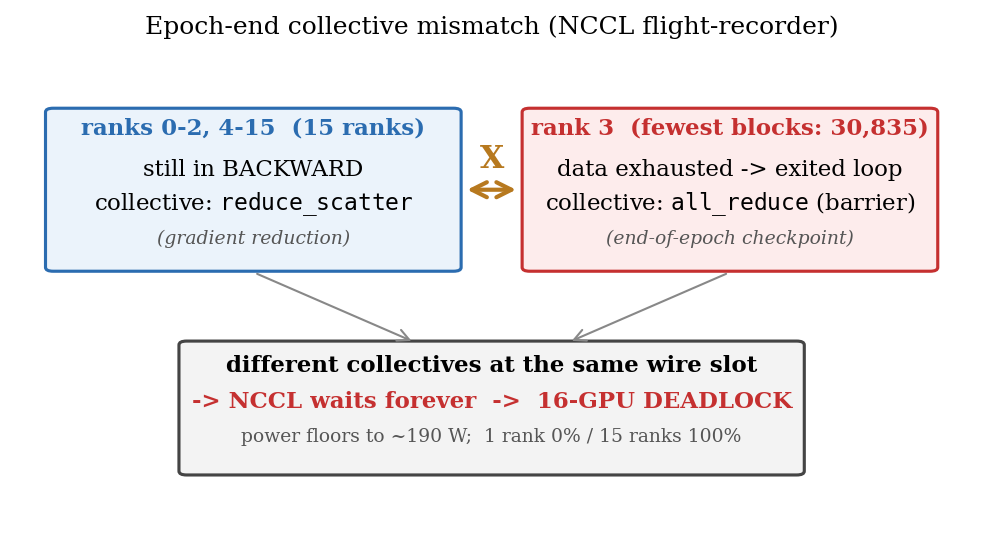}
\caption{Epoch-end collective mismatch. The shortest-data rank (rank 3, $30{,}835$
blocks) exits the loop one step early and enters the checkpoint \texttt{all\_reduce}
while the other fifteen ranks are still in the backward \texttt{reduce\_scatter}.
Different collectives at the same slot $\Rightarrow$ permanent NCCL deadlock.}
\label{fig:deadlock}
\end{figure}

\subsection{Root cause: per-rank packed-block imbalance}
The loader shards \emph{rows} evenly, but persona records vary in length, so the
per-rank token count varies and, after packing into 2048-token blocks, the block
counts differ:
\begin{center}
\begin{tabular}{lr}
\toprule
metric & value \\
\midrule
min blocks (rank 3) & 30{,}835 \\
max blocks (rank 11) & 30{,}888 \\
spread (max $-$ min) & 53 \\
\bottomrule
\end{tabular}
\end{center}
Rank 3 ran out 53 blocks (${\approx}7$ optimizer steps, at 8 blocks/step) before
rank 11, exited the loop, and walked into the save barrier while others reduced
gradients. The 53-block spread is \textbf{0.17\% of
the per-rank workload}---invisible to every metric except the one that mattered,
and, as a corollary, \textbf{invisible to smoke testing}: a \texttt{max\_steps=N}
run exercises only the loop body, where all ranks are symmetric, and exits before
the shortest rank exhausts its data. The hazard's probability is zero on every
prefix and one at the boundary.

\begin{figure}[t]
\centering
\includegraphics[width=0.92\linewidth]{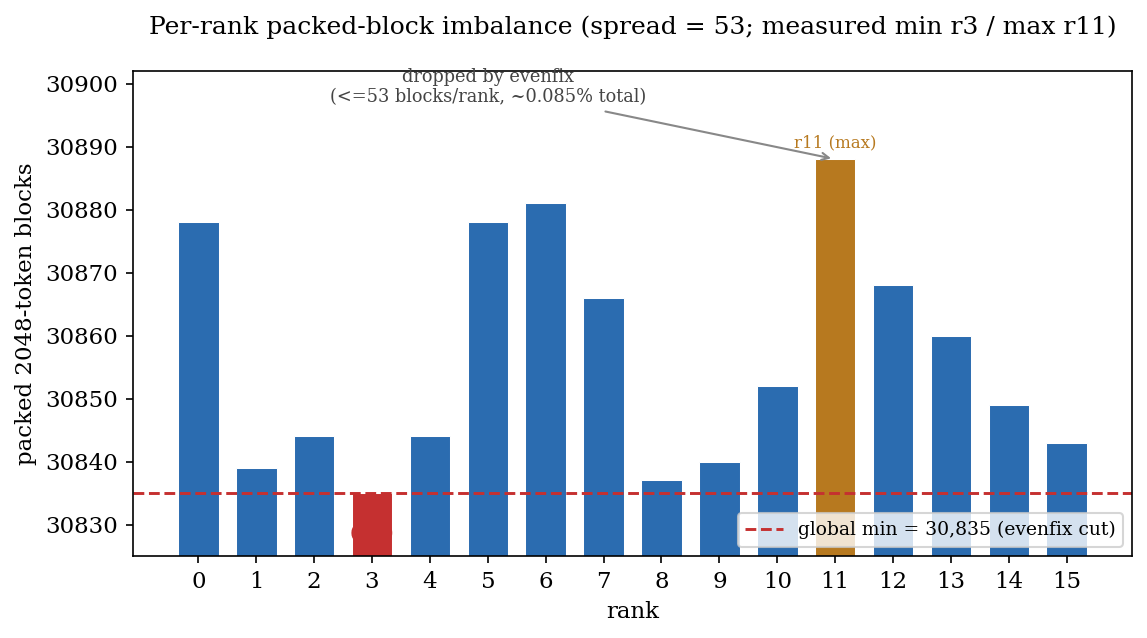}
\caption{Per-rank packed-block counts. Even row-sharding yields unequal block
counts (spread 53); the global minimum is the evenfix cut. The discarded tail is
${\le}53$ blocks/rank, ${\sim}0.085\%$ of tokens.}
\label{fig:blocks}
\end{figure}

\subsection{Prior art: this is \texttt{Join} / \texttt{drop\_last}, at the packing layer}
\label{sec:priorart}
The failure and remedy are documented PyTorch behavior; we are explicit rather than
imply discovery:
\begin{itemize}
  \item That \textbf{uneven per-rank inputs desynchronize collectives and hang the
  job} is the stated motivation for \texttt{DistributedDataParallel.join()} (since
  PyTorch 1.7, 2020) and the generic \texttt{torch.distributed.algorithms.Join}
  (1.10, 2021, covering ZeRO-style optimizers). The tutorial states it directly:
  ``if a rank has fewer inputs, then the other ranks will hang or error.''
  \item The standard remedy---equalize per-rank step counts by dropping the
  trailing remainder---is \texttt{DataLoader(drop\_last=True)} /
  \texttt{DistributedSampler}; HF Accelerate ships \texttt{even\_batches} and a
  \texttt{join\_uneven\_inputs} wrapper for exactly this.
  \item Per-rank step imbalance from \textbf{token packing} is itself an active
  topic (SlimPack; Hierarchical Balance Packing), framed there as efficiency.
\end{itemize}
We considered using \texttt{Join} directly to avoid dropping any data, but it is
\textbf{unavailable in our setting on two counts.} First, neither FSDP1
(\texttt{FullyShardedDataParallel}) nor FSDP2 (\texttt{fully\_shard}) implements
\texttt{Joinable}---only DDP and \texttt{ZeroRedundancyOptimizer} do---so an
FSDP-wrapped model cannot be passed into \texttt{with Join([...])}. Second,
\texttt{Join} shadows only the collectives issued \emph{inside a registered
Joinable's hooks} (DDP's gradient all-reduce, ZeRO's optimizer collectives), not
arbitrary user collectives; our deadlock was at a manual end-of-epoch checkpoint
\texttt{dist.barrier()}, outside that scope, which would desync even with
\texttt{Join} active. For FSDP with a manual save barrier the documented path is
exactly to \textbf{equalize per-rank step counts}---\texttt{drop\_last} upstream, or
an explicit \texttt{all\_reduce(MIN)} of the iteration count; we implement the
latter.\footnote{\url{https://docs.pytorch.org/tutorials/advanced/generic_join.html};
\url{https://docs.pytorch.org/docs/stable/distributed.algorithms.join.html}. Only
DDP and \texttt{ZeroRedundancyOptimizer} are \texttt{Joinable}; the FSDP sources
register no Join hook.} Note also that \texttt{drop\_last} only removes a rank's
\emph{partial} trailing batch; with micro-batch 1 over packed blocks there is no
partial batch, so \texttt{drop\_last} is inert and the cross-rank
\texttt{all\_reduce(MIN)} is what actually enforces equality.

Our ``evenfix'' is functionally the equalize-to-minimum pattern applied at the
\textbf{post-packing block layer} rather than the sampler---a placement choice, not
a new algorithm. We reframe the imbalance as a \emph{correctness} (deadlock) hazard
and contribute the worked B300 diagnosis and the preventive discipline of
\S\ref{sec:hardening}. That is the honest extent of the delta.

\subsection{evenfix --- the runtime guard we deployed}
Before the loop, each rank reduces its local block count to the global minimum:
\begin{lstlisting}[language=Python]
local_blocks = file_bytes // 4 // seq_len
min_blocks = dist.all_reduce(torch.tensor(local_blocks), op=ReduceOp.MIN)
# in the loop:
if blk_idx >= min_blocks:
    break   # every rank stops at exactly the same step
\end{lstlisting}
All ranks now run an identical number of iterations, so the terminal save-barrier
transition is symmetric and the deadlock is structurally impossible. The cost is
each rank's trailing $\text{local}-\text{min}$ blocks---at most 53 on any single
rank, \textbf{${\sim}0.085\%$ of all tokens in aggregate}---negligible. A live log
line confirms it:
\texttt{[evenfix] local=30,849 $\to$ min=30,835}. (A pad-to-maximum variant would
retain all data via duplicate blocks; we chose truncation for simplicity given the
negligible loss.)

\section{Calibrated Strong-Scaling Reference Numbers (B300)}
\label{sec:scaling}

We report $4/8/16$-GPU scaling as \textbf{reference data}, not a finding:
near-linear strong scaling for a compute-bound model at this size is the expected
regime, and competitive B300 figures already exist publicly (e.g.\ MLPerf Training
v5.1/v6.0, \S\ref{sec:related}). What is useful is the \emph{calibrated absolute
numbers}. Each point is a \textbf{complete measured epoch} (4 GPUs $15{,}420$
steps, 8 GPUs $7{,}710$, 16 GPUs $3{,}854$; medians over $n=1{,}538/767/385$).

\begin{center}
\begin{tabular}{rrrrrrr}
\toprule
GPUs & Nodes & tok/s & tok/s/GPU & 1 epoch (h) & GPU$\cdot$h & Strong-scaling eff. \\
\midrule
4 & 0.5 & 13.4k & 3{,}350 & 20.9 & 83.7 & 100\% \\
8 & 1 & 26.7k & 3{,}338 & 10.5 & 84.1 & 100\% \\
16 & 2 (IB) & 53.0k & 3{,}313 & 5.3 & 84.7 & 99\% \\
\bottomrule
\end{tabular}
\end{center}

Per-GPU throughput is \textbf{flat within ${\sim}1\%$} ($3{,}313$--$3{,}350$) and
GPU$\cdot$h per epoch is conserved at \textbf{${\sim}84$} (4-GPU baseline; the only
loss is a ${\sim}1\%$ tax at 16 GPUs from the two-node IB boundary---4/8 GPUs are
NVLink-only). Final loss was preserved across all three ($0.674/0.673/0.67$),
confirming scaling does not perturb convergence. The conservation is the positive
control for the \S\ref{sec:triage} taxonomy: GPUs sat at ${\sim}940$--$1040$~W
(compute regime) throughout, confirming communication was hidden behind compute.

\paragraph{Implication for in-network aggregation (SHARP).} The same ${\sim}1\%$
two-node tax bounds the upside of switch-based collective offload. NVIDIA SHARP
performs the gradient reduction inside the InfiniBand fabric, attacking
\emph{communication} cost; but at 16 GPUs the reduction is already overlapped
behind backward compute ($99\%$ strong-scaling efficiency, GPUs at $940$--$1040$~W),
so the \emph{exposed} communication SHARP could remove is at most the ${\sim}1\%$
scaling gap---and only the communication fraction of that. By Amdahl's law the
throughput ceiling is therefore ${<}1\%$ in this regime. We did not provision SHARP
on this fabric (the aggregation manager grants a zero resource quota---%
\texttt{osts:0, max\_groups:0}---and \texttt{sharp\_coll\_test} blocks at group
allocation), and the expected gain is correspondingly marginal here. This is a
property of the operating point, not of SHARP: its value grows at larger scale,
where ring/tree all-reduce latency becomes exposed and the $O(\log N)$ in-network
tree dominates. Our reference numbers should not be read as evidence against SHARP
there.

\begin{figure}[t]
\centering
\includegraphics[width=0.78\linewidth]{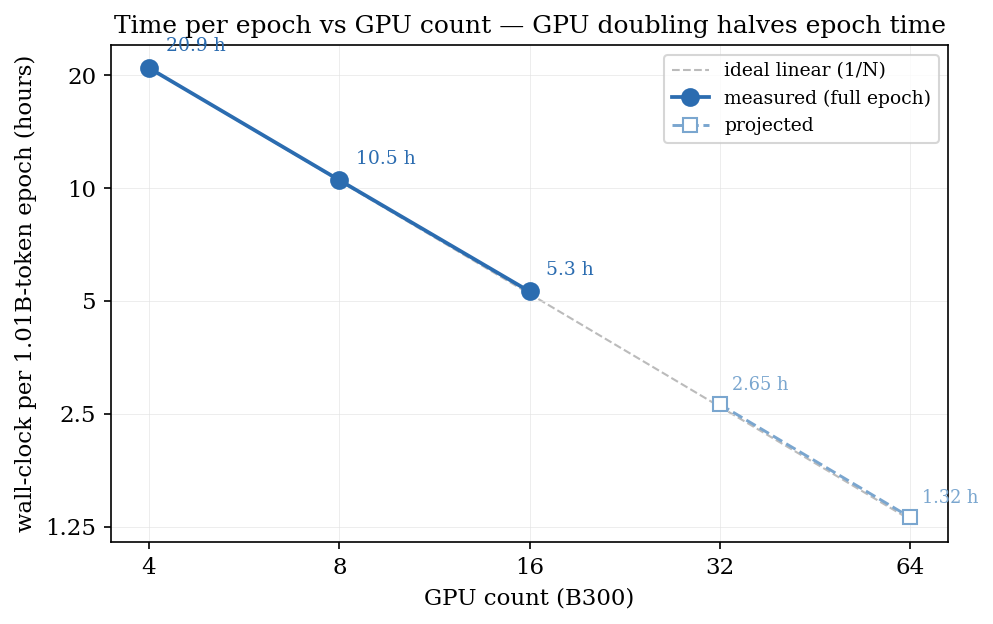}
\caption{Wall-clock per 1.01B-token epoch vs.\ GPU count. Measured at $4/8/16$;
$32/64$ projected. GPU doubling halves epoch time.}
\label{fig:tpe}
\end{figure}

\begin{figure}[t]
\centering
\includegraphics[width=0.78\linewidth]{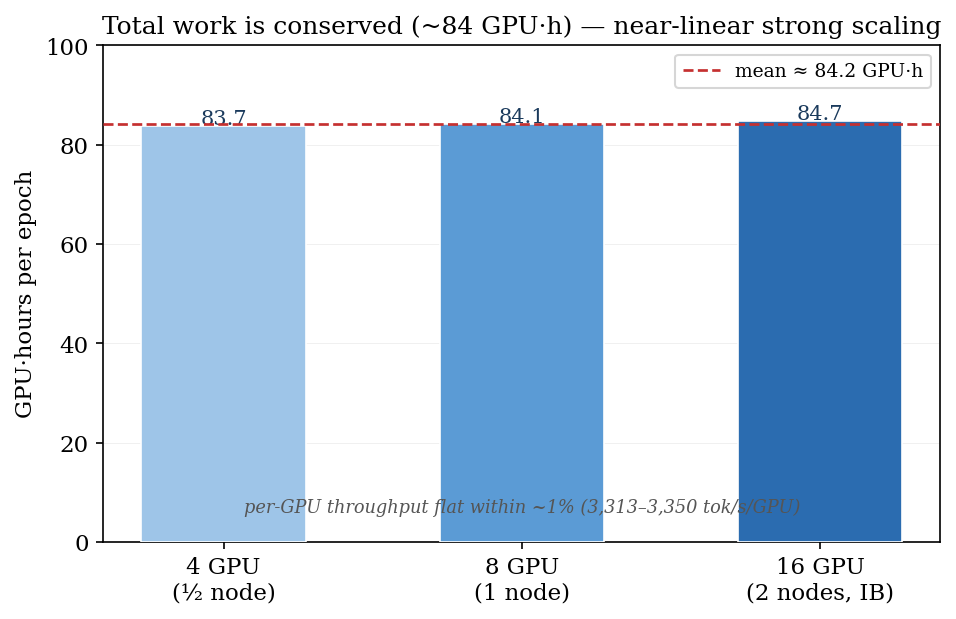}
\caption{GPU$\cdot$hours per epoch conserved at ${\sim}84$ across $4/8/16$ GPUs
($99$--$100\%$ efficiency, 4-GPU baseline): total work preserved, no communication
bottleneck across the two-node IB boundary.}
\label{fig:gpuh}
\end{figure}

The clean from-scratch epoch that \textbf{validated the \S\ref{sec:case} fix} is
the 16-GPU row: 3{,}854 steps, rc $=0$ on the first attempt, final loss 0.67,
median 53.0k tok/s. The preflight gate passed (\texttt{min\_blocks $=30{,}835$}),
evenfix logged \texttt{local=30,849 $\to$ min=30,835}, and the epoch-end save
completed with \textbf{zero watchdog, abort, or deadlock traces}---the transition
that previously hung for three hours now passes cleanly. The 62~GB checkpoint loads
and generates on-format, fluent in-domain text.

\section{Operational Hardening and Cost Accounting}
\label{sec:hardening}

\subsection{The cost of one silent failure}
A single occurrence of the \S\ref{sec:case} deadlock cost, conservatively:
\begin{center}
\begin{tabular}{lr}
\toprule
Component & GPU$\cdot$h \\
\midrule
The hang itself (16 GPU $\times$ ${\sim}3$~h before watchdog) & ${\sim}48$ \\
Unrecoverable progress (pre-fix epoch-end-only save; ${\sim}5$~h $\times$ 16 GPU) & ${\sim}85$ \\
\textbf{Total per incident} & \textbf{${\sim}130$} \\
\bottomrule
\end{tabular}
\end{center}
That is \textbf{${\sim}1.5$ epochs of compute} (${\sim}84$~GPU$\cdot$h/epoch) thrown
away by a sub-percent, sub-visible imbalance. Separately, the \S\ref{sec:contention}
starvation incident cost an estimated \textbf{${\sim}+3$~h wall-clock} on a 5.3~h
epoch (${\approx}+50$~GPU$\cdot$h) before diagnosis---now caught in a single
\texttt{nvidia-smi} sample.

\subsection{The preflight gate --- block the launch, not the run}
Because ``smoke pass $\neq$ full safe,'' we verify \emph{before} the expensive
launch. A gate checks six invariants and refuses to start an unsafe run; its pass
log doubles as a recorded clean-run pre-validation artifact. \textbf{Measured
wall-clock to run all six checks across both nodes: 2.71~s.}

\begin{center}
\begin{tabular}{cll}
\toprule
\# & Check & Failure it prevents \\
\midrule
1 & Cache integrity $+$ per-rank block balance & missing shards; records evenfix precond. \\
2 & Token-total sanity (${\sim}1.01$B $\pm10\%$) & partial cache build \\
3 & GPU occupancy $=0$ on both nodes & zombie/competing processes \\
4 & \texttt{master\_port} free & stale rendezvous \\
5 & Checkpoint disk headroom (${\geq}150$~GB) & save-stage failure \\
6 & (optional) IB bus-bandwidth probe & fabric degradation (opt-in) \\
\bottomrule
\end{tabular}
\end{center}

A nonzero block spread is \textbf{WARN, not FAIL} (evenfix handles it); the gate
records \texttt{global\_min\_blocks}. It is \textbf{hard-wired into the launcher},
so every full run is admitted only after passing. The economics are stark: a
\textbf{2.71-second} check stands between submission and a failure that, when it
slips through, costs \textbf{${\sim}130$~GPU$\cdot$h}.

\subsection{The external watcher --- observe from outside the run}
Nodes that deadlock cannot report their own death. A \textbf{session-independent
watcher} on a separate host detects completion (epoch checkpoint appears), crash
(process gone, no checkpoint), \textbf{hang} (step counter stalls, or the
\S\ref{sec:triage} power/utilization deadlock signature persists across two
samples), and throughput floor---then emits an out-of-band alert. Recoverability
was addressed orthogonally with periodic checkpointing
(\texttt{save\_every=500}, ${\sim}42$~min), weights-only resume, an increased
watchdog timeout ($10 \to 30$~min), and launcher auto-retry---standard practice
mentioned for completeness.

\section{Related Work}
\label{sec:related}

\textbf{Uneven inputs and packing balance.} That uneven per-rank inputs cause
collective desync and hangs is the documented rationale for PyTorch's
\texttt{DistributedDataParallel.join()} (1.7) and the generic \texttt{Join} (1.10),
and is mitigated by \texttt{DataLoader(drop\_last=True)} / \texttt{DistributedSampler}
and HF Accelerate's \texttt{even\_batches} / \texttt{join\_uneven\_inputs}. Notably,
\texttt{Join} is implemented only for DDP and ZeRO, \textbf{not FSDP}
(\S\ref{sec:priorart}), so FSDP users with uneven per-rank inputs must equalize step
counts manually. Per-rank
imbalance from token packing is active research (SlimPack, arXiv:2509.26246;
Hierarchical Balance Packing, arXiv:2503.07680), framed as efficiency; we reframe it
as a correctness hazard but claim no algorithmic novelty.

\textbf{Telemetry-based stall diagnosis.} Using GPU power and hardware telemetry to
detect stalls/stragglers---including the ``100\% utilization during an NCCL
hang'' symptom---is established in vendor/SRE practice (NVIDIA DCGM; Megatron/NeMo
straggler detection; troubleshooting playbooks) and in research on telemetry-based
anomaly detection (arXiv:2510.26008; Mycroft, arXiv:2509.03018). Concurrent 2026
work extends telemetry-based failure detection further: observability-aware early
warning that models monitoring-pipeline degradation itself (arXiv:2603.28781), and
performance-counter-based stress estimation beyond utilization\%
(arXiv:2511.05067). Phase-resolved
power, including a named ``execution-idle'' state, has been characterized
(arXiv:2604.04745; ASPLOS'24 LLM power characterization). We contribute
B300-calibrated absolute bands, not a new principle.

\textbf{Scaling/reliability at scale.} Large-scale training experience reports---the
OPT-175B logbook, BLOOM, MegaScale, the Llama 3 herd reliability section, Imbue's
bare-metal-to-70B account---document emergent failures at thousands of GPUs. Ours is
modest scale (16 GPUs) with a documented root cause; we position it as a focused
field note. Closest in genre is a concurrent operational report on a 504-GPU
\emph{B200} production cluster (arXiv:2605.09370), which analyzes pre-training
failures via a multi-layer Prometheus metric suite ($\sim$751 metrics at 30\,s
granularity) and recovery workflow; it does not map board power to execution phase,
and its workload is MoE pre-training rather than dense full fine-tuning. Our
contributions are therefore complementary: single-command power-band triage that
needs no metric infrastructure, and a full-fine-tuning-specific packing-imbalance
hazard. Public B300 scaling references include MLPerf Training v5.1/v6.0.

\textbf{Fault tolerance.} ZeRO/FSDP sharding, overlapped communication, and
checkpoint/restart are well established; our use is standard.

\section{Threats to Validity and Scope}
\label{sec:threats}
\begin{itemize}
  \item \textbf{Scale.} We measured to 16 GPUs / 2 nodes. The deadlock mechanism is
  scale-independent, but the efficiency numbers (\S\ref{sec:scaling}) should not be
  extrapolated past the two-node IB boundary, and at-scale reliability lessons do
  not follow from a 16-GPU study.
  \item \textbf{Single model/dataset.} The generality argument is structural, not
  empirical across models.
  \item \textbf{Power thresholds are hardware/firmware-specific.} The
  \S\ref{sec:triage} wattages are B300 values; the taxonomy and shape
  discriminators should transfer, absolute thresholds must be re-calibrated.
  \item \textbf{Data-vs-RAM assumption.} The \S\ref{sec:negative} conclusion holds
  only while the dataset fits in page cache; it inverts for datasets $>$ RAM.
  \item \textbf{Reconstructed cost.} The \S\ref{sec:hardening} starvation cost is
  estimated from logs, not a controlled measurement.
\end{itemize}

\section{Conclusion and Artifact Availability}
We shared operational experience full-fine-tuning Qwen3-32B on 16 $\times$ B300.
Rather than a new algorithm, we offer four practitioner artifacts: a
B300-calibrated power-draw triage procedure (utilization lies; power tells the
truth); honest negative results showing local-cache optimization buys no throughput
when the dataset fits in RAM and that an earlier ``storage collapse'' was really
CPU/NFS contention; calibrated $4/8/16$-GPU strong-scaling reference numbers; and a
worked epoch-end deadlock---explicitly the documented \texttt{Join}/\texttt{drop\_last}
failure, applied at the packing layer---hardened by a 2.71-second preflight gate and
an external watcher that convert a ${\sim}130$~GPU$\cdot$h silent failure into an
instant rejection. The transferable takeaways are operational: \emph{watch power,
not utilization}, and \emph{verify invariants before launch---a passing smoke test
is not evidence of a safe full run.}

\paragraph{Artifacts.} The FSDP trainer (with evenfix), the preflight gate, the
external watcher, the pretokenizer/packer, and the scaling/benchmark scripts are
available from the authors upon reasonable request.

\appendix
\section{Reproduction sketch}
\begin{enumerate}
  \item Pretokenize and pack the corpus into per-rank \texttt{.i32} block files
  (\texttt{pretokenize\_persona.py}); record per-rank block counts.
  \item Run the gate (\texttt{preflight\_gate.sh}, \texttt{WORLD=16 NPROC=8
  NODES=2}); confirm checks 1--5 pass (${\sim}3$~s) and note \texttt{global\_min\_blocks}.
  \item Launch (\texttt{full\_ft\_launch.sh}); the launcher invokes the gate, then
  torchrun across both nodes with evenfix active.
  \item Monitor with the one-line power query (\S\ref{sec:triageproc}) and/or the
  external watcher; expect ${\sim}940$--$1040$~W compute and a clean ${\sim}190$~W
  rank-0-only checkpoint at epoch end.
\end{enumerate}

\section{Key measured quantities}
\begin{center}\small
\begin{tabular}{ll}
\toprule
Quantity & Value \\
\midrule
Model & Qwen3-32B, 32.76B dense \\
World & 16 GPU (2 $\times$ 8 B300), FSDP FULL\_SHARD / ZeRO-3, bf16 \\
Tokens/step & 262{,}144 \\
Steps/epoch & ${\sim}3{,}854$ \\
Tokens/epoch & ${\sim}1.01$B (after packing) \\
Per-rank blocks & min 30{,}835 / max 30{,}888 / spread 53 \\
Tokens dropped by evenfix & ${\sim}0.085\%$ \\
Throughput (16/8/4 GPU) & 53.0k / 26.7k / 13.4k tok/s \\
Memory & 82~GB/GPU (grad-ckpt off), 30\% of 275~GB \\
Loss & $1.60 \to 0.67$ (1 epoch) \\
Epoch wall-clock (16 GPU) & ${\sim}5.3$~h \\
GPU$\cdot$h/epoch & ${\sim}84$ (conserved across 4/8/16) \\
Data A/B (NFS cold / warm / local) & 52.8k / 53.0k / 53.0k tok/s \\
Cold buffered storage read & 1.31 GB/s \\
Preflight gate wall-clock & 2.71~s (6 checks, 2 nodes) \\
Hang duration (pre-fix) & ${\sim}3$~h before watchdog \\
Cost per silent failure & ${\sim}130$~GPU$\cdot$h (${\sim}1.5$ epochs) \\
\bottomrule
\end{tabular}
\end{center}

\end{document}